\documentclass{article}

\usepackage{PRIMEarxiv}

\usepackage[utf8]{inputenc} 
\usepackage[T1]{fontenc}    
\usepackage{url}            
\usepackage{booktabs}       
\usepackage{amsfonts}       
\usepackage{nicefrac}       
\usepackage{microtype}      
\usepackage{lipsum}
\usepackage{fancyhdr}       
\usepackage{graphicx}       
\usepackage{arabtex}
\usepackage{arabtex}
\usepackage{utf8}
\usepackage{amsmath}
\usepackage{algorithm}
\usepackage{algpseudocode}
\usepackage{natbib}
\usepackage[colorlinks=true,
linkcolor=blue,
citecolor=blue,
urlcolor=blue]{hyperref}
\setcode{utf8}
\usepackage{dblfloatfix}
\usepackage{tabularx} 
\usepackage{longtable}

\graphicspath{{media/}}     

\title{A Dirichlet-Quantile Bootstrap approach for the probabilistic resampling of time series}

\author{
  Ahmed Hamimes \\
  BIOSTIM Laboratory, Medicine Faculty \\
  Salah Boubnider Constantine 03 University \\
  Constantine, Algeria\\
  \texttt{ahmed.hamimes@univ-constantine3.dz} \\
   \And
  Abdennour Boulesnane \\
BIOSTIM Laboratory, Medicine Faculty \\
Salah Boubnider Constantine 03 University \\
Constantine, Algeria\\
\texttt{aboulesnane@univ-constantine3.dz} \\
}
\renewenvironment{abstract}
{
  \small    
  \begin{center}
  \bfseries Abstract  
  \end{center}
  \begin{quote}       
}
{
  \end{quote}
}

\begin{document}
\maketitle

\begin{abstract}
This work presents DPQBootstrap, a probabilistic resampling method for time series based on a Dirichlet quantile bootstrap and rank-based temporal reconstruction. The method aims to generate pseudo-series that preserve the main characteristics of the original series, including variability, quantiles, extreme values, and part of its temporal dependence. A simulation study compares DPQBootstrap with the Maximum Entropy Bootstrap implemented in the meboot package across eight simulation scenarios. The results show that DPQBootstrap achieves 106 wins over 86 for meboot according to the Interval Score, indicating a strong ability to preserve the statistical properties of the original time series. In addition, an empirical application was conducted within a Bayesian ARMA forecasting framework, with and without DPQBootstrap integration. The results indicate that incorporating DPQBootstrap can improve point forecast accuracy and the quality of predictive intervals, particularly when combined with appropriately specified ARMA models.
\end{abstract}

\keywords{Time series \and Bootstrap \and Resampling \and Dirichlet distribution \and Quantile resampling \and Maximum Entropy Bootstrap \and Bayesian ARMA forecasting.}

\section{Introduction}
Time series analysis occupies a central role in numerous applied fields, including economics, hydrology, energy, environmental studies, and management sciences \cite{box2015time}. In such contexts, observed data are rarely independent and may exhibit trends, seasonality, varying variance, extreme values, or temporal dependence. These characteristics make statistical inference more challenging, as classical methods based on independence assumptions may lead to an inadequate assessment of uncertainty \cite{brockwell2002introduction}.

The bootstrap, introduced by \cite{Efron1992}, is one of the most widely used approaches for approximating the distribution of a statistic from observed data. Its principle is to generate pseudo-samples to estimate the uncertainty associated with a statistic of interest. However, the classical bootstrap based on sampling with replacement is primarily designed for independent and identically distributed observations. When applied to time series, such an approach may disrupt temporal ordering and weaken the data's dependence structure. Consequently, several extensions have been proposed to accommodate dependent data, including the block bootstrap of \cite{kunsch1989jackknife} and the stationary bootstrap of \cite{politis1994stationary}.

Within the context of time series, an important contribution is the Maximum Entropy Bootstrap proposed by \cite{Vinod2009} and implemented in the meboot R package. This method aims to generate pseudo-time series while respecting certain properties of the original series, based on a maximum entropy principle. It has become an important reference for time-series resampling, particularly when the objective is to preserve temporal structure without imposing a strict parametric model.\\
In parallel, Bayesian resampling methods have also been developed, notably through \cite{rubin1981bayesian} Bayesian bootstrap. This approach replaces classical empirical resampling with random observation weighting using Dirichlet-distributed weights. This idea introduces probabilistic uncertainty into the observed empirical distribution. Dirichlet distributions and Dirichlet processes, notably studied by \cite{ferguson1973bayesian}, therefore provide a natural framework for representing uncertainty about an unknown distribution.

In this work, we present DPQBootstrap, an R package dedicated to probabilistic resampling for time series. The proposed method is based on a Dirichlet quantile bootstrap combined with rank-preserving temporal reconstruction. The main idea is to construct a continuous bootstrap distribution from the ordered observations, assign random Dirichlet weights to the associated intervals, and then generate pseudo-observations via an inverse-quantile transformation. The resulting pseudo-series is subsequently reconstructed from the original ranks to preserve the relative structure of the original series. The proposed methodology therefore lies at the intersection of quantile bootstrap methods, the Bayesian bootstrap, and time series resampling.

The objective of this article is twofold. First, we present the methodological foundations of DPQBootstrap and explain how it differs from the Maximum Entropy Bootstrap. Second, a simulation study is conducted to evaluate its ability to preserve essential characteristics of time series, including the mean, median, variability, quantiles, extreme values, and first-order autocorrelation. The comparison is performed against the meboot package using the Interval Score, a criterion that jointly accounts for the width of bootstrap intervals and their ability to contain the reference statistic.\\  
Therefore, this work does not seek to replace parametric forecasting models such as ARMA or SARIMA. Rather, it proposes a complementary probabilistic resampling method for time series, particularly for complex series exhibiting irregular structures, non-Gaussian distributions, extreme values, or limited sample sizes.

\section{Background and Methodological Foundations}

\subsection{Time Series Bootstrap and Pseudo-Series Reconstruction} 
The proposed method is situated within the framework of time-series resampling, which aims to generate pseudo-series that retain key characteristics of the original series. Unlike the classical bootstrap, which directly resamples observations without accounting for their temporal organization, time series resampling seeks to generate artificial series while preserving, as far as possible, the relative structure of the data over time.\\
Let an observed time series be given by :
\begin{equation*} X = (x_1, x_2, \ldots, x_n). \end{equation*}  
The observations are first arranged in increasing order:  \begin{equation*} x_{(1)} \leq x_{(2)} \leq \cdots \leq x_{(n)}. \end{equation*} 
For each observation $x_t$, its rank in the ordered series is defined as:
\begin{equation*} 
	r_t = \operatorname{rank}(x_t), \qquad t = 1, \ldots, n.
\end{equation*}

This rank indicates the relative position of the observation within the empirical distribution of the series. For example, a small value receives a low rank, whereas a large value receives a high rank.\\
The main idea is then to generate new pseudo-observations from a bootstrap distribution and subsequently place them back into the series according to the original ranks. Thus, the resulting pseudo-series is reconstructed as:  \begin{equation*} X_t^* = x_{(r_t)}^*, \qquad t = 1, \ldots, n. \end{equation*}
Also, if the original observation $x_t$ occupied rank $r_t$, the bootstrap value at the same time point, $x_t^*$, is assigned the pseudo-observation corresponding to the same rank. This reconstruction step preserves the relative structure of the original series, in particular the temporal locations of low, intermediate, and high values.

The time series resampling procedure used here is therefore not limited to reproducing the marginal distribution of the data. It also seeks to retain part of the series' temporal organization through rank-based reconstruction. This approach is particularly useful for generating pseudo-series for the study of uncertainty, for constructing bootstrap intervals, or for analyzing the stability of statistics derived from a time series.

\subsection{The Maximum Entropy Bootstrap Method}

The method is based on the idea of constructing a bootstrap density that is as uninformative as possible while remaining compatible with the constraints imposed by the observed data. This approach is based on the maximum entropy principle: among all admissible densities, the one that maximizes Shannon entropy is selected. The target density is therefore defined as:
\begin{equation*}
	f^*=\arg\max_{f in\mathcal{F}}
	\left\{
	-\int f(x)\ln f(x)\,dx
	\right\}.
\end{equation*}

The objective of this maximization is to obtain a density that does not introduce additional artificial information. In other words, the density $f^*$ is selected to be as neutral as possible with respect to the available data.

Once the density $f^*$ has been obtained, its associated cumulative distribution function is constructed as:
\begin{equation*}
	F^*(x)
	=
	\int_{-\infty}^{x} f^*(s)\,ds.
\end{equation*}

Pseudo-observations are then generated through the inverse quantile transformation:
\begin{equation*}
	x_i^*
	=(F^*)^{-1}(u_i), u_i \sim U(0,1).
\end{equation*}

This step makes it possible to generate new values from the bootstrap distribution constructed according to the maximum entropy principle.

Finally, to obtain a pseudo-time series, the pseudo-observations are assigned to their temporal positions according to the original ranks of the observed series. If $r_t$ denotes the rank of observation $x_t$, the pseudo-series is reconstructed as:
\begin{equation*}
	X_t^*
	=
	x_{(r_t)}^*.
\end{equation*}

Thus, the method is not limited to generating new values. It also seeks to preserve the relative structure of the original series by retaining the temporal organization of the ranks.

The Maximum Entropy Bootstrap is based on constructing a density by entropy maximization. This density is selected to be as uninformative as possible while remaining compatible with the observed data. However, this approach generally leads to a unique density. In other words, once $f^*$ has been constructed, the generating distribution used to produce the pseudo-observations is fixed.

In the Bayesian approach, the objective is different. Rather than seeking a single optimal density, the aim is to represent the uncertainty associated with the data's unknown distribution. Indeed, the observed series constitutes only a limited sample from the underlying process. Consequently, there is empirical uncertainty regarding the probability mass associated with each observation in the generating distribution. To account for this uncertainty, random Dirichlet weights are introduced:
\begin{equation*}
	(W_1,\ldots,W_n)
	\sim
	\operatorname{Dirichlet}
	(\alpha_1,\ldots,\alpha_n).
\end{equation*}

Each weight represents the probability mass assigned to an observation or to an interval constructed around an ordered observation. Thus, unlike the Maximum Entropy Bootstrap, in which the probability masses may be determined through a maximum-entropy construction, the Bayesian approach allows these masses to vary from one bootstrap replication to another. This idea makes it possible to replace the maximum-entropy density 
\section{Proposed Method: DPQBootstrap}
\subsection{Dirichlet Mixture Quantile Bootstrap Based on Stick-Breaking}

In the proposed approach, the objective is not to construct the density that maximizes entropy directly. Instead, this density is replaced by a flexible Bayesian density estimated using a finite mixture of normal distributions:
\begin{equation*}
	f_D(x)
	=
	\sum_{j=1}^{C}
	\pi_j
	\mathcal{N}(x \mid \mu_j,\sigma_j^2).
\end{equation*}

Here, $C$ denotes the maximum number of mixture components, $\pi_j$ the weight of component $j$, $\mu_j$ its mean, and $\sigma_j^2$ its variance.

This density makes it possible to approximate non-Gaussian, asymmetric, or multimodal distributions, as well as distributions affected by extreme observations. It therefore provides a more flexible alternative to a single normal distribution and is more compatible with a Bayesian formulation.

The weighting of the mixture is constructed using a finite stick-breaking representation. Latent variables are introduced as:
\begin{equation*}
	q_j
	\sim
	\operatorname{Beta}(1,\alpha),
	\qquad
	j=1,\ldots,C.
\end{equation*}

The first raw weight is defined as:
\begin{equation*}
	w_1=q_1.
\end{equation*}

For the subsequent components, we define:
\begin{equation*}
	w_j
	=
	q_j
	\prod_{l=1}^{j-1}(1-q_l),
	\qquad
	j=2,\ldots,C.
\end{equation*}

Since the number of components is finite, the raw weights are subsequently normalized:
\begin{equation*}
	\pi_j
	=
	\frac{w_j}
	{\sum_{l=1}^{C}w_l},
	\qquad
	j=1,\ldots,C.
\end{equation*}

Thus,
\begin{equation*}
	\pi_j\geq0,
	\qquad
	\sum_{j=1}^{C}\pi_j=1.
\end{equation*}

The parameter $\alpha$ controls the concentration of the mixture. When $\alpha$ is small, the probability mass may be concentrated on a small number of dominant components. When $\alpha$ is larger, the weights tend to be distributed over a larger number of components.

Starting from the density $f_D$, the cumulative distribution function is defined as:
\begin{equation*}
	F_D(x)
	=
	\int_{-\infty}^{x}f_D(s)\,ds.
\end{equation*}

Since $f_D$ is a mixture of normal distributions, this yields:
\begin{equation*}
	F_D(x)
	=
	\sum_{j=1}^{C}
	\pi_j
	\Phi
	\left(
	\frac{x-\mu_j}{\sigma_j}
	\right),
\end{equation*}
where $\Phi$ denotes the cumulative distribution function of the standard normal distribution.

The function $F_D$ plays the same role as the cumulative distribution function derived from the maximum-entropy density in the MEB. However, in the present approach, it is derived from a Bayesian mixture model rather than from a principle of entropy maximization.

Pseudo-observations are generated from $F_D$ through the inverse quantile transformation:
\begin{equation*}
	x_i^*
	=
	F_D^{-1}(u_i),
	\qquad
	u_i\sim U(0,1).
\end{equation*}

In practice, the inverse does not have a simple analytical form. It is therefore approximated numerically on an ordered grid:
\begin{equation*}
	g_1<g_2<\cdots<g_M.
\end{equation*}

The corresponding values are computed as:
\begin{equation*}
	F_D(g_1),F_D(g_2),\ldots,F_D(g_M).
\end{equation*}

For each $u_i$, an index $k$ is identified such that:
\begin{equation*}
	F_D(g_{k-1})
	<
	u_i
	\leq
	F_D(g_k).
\end{equation*}

The pseudo-observation is then approximated by linear interpolation:
\begin{equation*}
	x_i^*
	=
	g_{k-1}
	+
	\frac{
		u_i-F_D(g_{k-1})
	}{
		F_D(g_k)-F_D(g_{k-1})
	}
	(g_k-g_{k-1}).
\end{equation*}

This step produces continuous pseudo-values generated from the estimated Bayesian density. To obtain an ordered sequence of pseudo-observations, one may use ordered uniform draws or increasing quantile windows associated with the ranks $r_t$.

The pseudo-values generated by inverse quantile transformation correspond to pseudo-observations associated with ranks. To reconstruct a pseudo-time series, the original ranks $r_t$ are used. The reconstruction is given by:
\begin{equation*}
	x_t^*
	=
	x_{(r_t)}^*,
	\qquad
	t=1,\ldots,n.
\end{equation*}

The proposed method can be summarized by the following scheme:
\begin{equation*}
	X=(x_1,\ldots,x_n),
\end{equation*}

\begin{equation*}
	x_{(1)}\leq\cdots\leq x_{(n)},
\end{equation*}

\begin{equation*}
	r_t=\operatorname{rank}(x_t),
\end{equation*}

\begin{equation*}
	q_j\sim\operatorname{Beta}(1,\alpha),
\end{equation*}

\begin{equation*}
	w_1=q_1,
	\qquad
	w_j=q_j\prod_{l=1}^{j-1}(1-q_l),
\end{equation*}

\begin{equation*}
	\pi_j=
	\frac{w_j}
	{\sum_{l=1}^{C}w_l},
\end{equation*}

\begin{equation*}
	f_D(x)
	=
	\sum_{j=1}^{C}
	\pi_j
	\mathcal{N}(x\mid\mu_j,\sigma_j^2),
\end{equation*}

\begin{equation*}
	F_D(x)
	=
	\sum_{j=1}^{C}
	\pi_j
	\Phi
	\left(
	\frac{x-\mu_j}{\sigma_j}
	\right),
\end{equation*}

\begin{equation*}
	x_i^*
	=
	F_D^{-1}(u_i),
	\qquad
	u_i\sim U(0,1),
\end{equation*}

\begin{equation*}
	X_t^*
	=
	x_{(r_t)}^*,
\end{equation*}

\begin{equation*}
	T^{*(b)}
	=
	T\left(X^{*(b)}\right).
\end{equation*}

\subsection{Computational Method for the Dirichlet-Based Quantile Bootstrap}

The first method, based on a Dirichlet mixture with stick-breaking weighting, provides considerable flexibility for approximating the generating distribution. However, it remains more complex to estimate and implement. For this reason, we retain in the DPQBootstrap package a second, more direct formulation based on Dirichlet weights assigned to intervals constructed around the ordered observations.

Starting from the ordered observations, a partition of the real line is constructed using intermediate points:
\begin{equation*}
	z_0,z_1,\ldots,z_n.
\end{equation*}

The internal boundaries are defined as the midpoints between two consecutive observations:
\begin{equation*}
	z_i
	=
	\frac{x_{(i)}+x_{(i+1)}}{2},
	\qquad
	i=1,\ldots,n-1.
\end{equation*}

For the boundaries at the extremes, a simple extrapolation is used:
\begin{equation*}
	z_0
	=
	x_{(1)}
	-
	\frac{x_{(2)}-x_{(1)}}{2},
	\qquad
	z_n
	=
	x_{(n)}
	+
	\frac{x_{(n)}-x_{(n-1)}}{2}.
\end{equation*}

Thus, each ordered observation $x_{(i)}$ is associated with an interval:
\begin{equation*}
	I_i
	=
	[z_{i-1},z_i],
	\qquad
	i=1,\ldots,n.
\end{equation*}

The width of this interval is:
\begin{equation*}
	\Delta_i
	=
	z_i-z_{i-1}.
\end{equation*}

This construction transforms discrete observations into continuous support, thereby enabling quantile-based bootstrap generation.

Instead of assigning each interval a fixed probability equal to $1/n$, the proposed method introduces probabilistic uncertainty into these probability masses. A vector of weights is drawn:
\begin{equation*}
	(W_1,W_2,\ldots,W_n)
	\sim
	\operatorname{Dirichlet}
	(\alpha,\alpha,\ldots,\alpha),
\end{equation*}
where $\alpha>0$ is a concentration parameter.

The weights satisfy:
\begin{equation*}
	W_i\geq0,
	\qquad
	\sum_{i=1}^{n}W_i=1.
\end{equation*}

Each weight $W_i$ represents the probability mass assigned to interval $I_i$. When $\alpha=1$, a uniformly random weighting of the Bayesian bootstrap type is obtained. As $\alpha$ increases, the weights become more homogeneous. As $\alpha$ decreases, the weights become more variable, thereby increasing the dispersion of the generated pseudo-series.

Conditional on the Dirichlet weights, a piecewise density is defined over the intervals $I_i$. For $x\in I_i$, we define:
\begin{equation*}
	f_D(x)
	=
	\frac{W_i}{\Delta_i}.
\end{equation*}

The complete density can be written as:
\begin{equation*}
	f_D(x)
	=
	\sum_{i=1}^{n}
	\frac{W_i}{\Delta_i}
	\mathbf{1}_{[z_{i-1},z_i]}(x).
\end{equation*}

This density is a random and continuous version of the empirical distribution. It retains the information contained in the ordered observations while introducing Dirichlet-type uncertainty into the probability masses.

Unlike the Maximum Entropy Bootstrap, the objective is therefore not to seek a density $f^*$ satisfying:
\begin{equation*}
	f^*
	=
	\arg\max_f
	\left[
	-
	\int f(x)\ln f(x)\,dx
	\right].
\end{equation*}

Instead, the density used is a Bayesian bootstrap density:
\begin{equation*}
	f_D(x)
	=
	\sum_{i=1}^{n}
	\frac{W_i}{z_i-z_{i-1}}
	\mathbf{1}_{[z_{i-1},z_i]}(x),
	\qquad
	W
	\sim
	\operatorname{Dirichlet}
	(\alpha,\ldots,\alpha).
\end{equation*}

This formulation avoids the maximum-entropy optimization step while retaining continuous generation of pseudo-observations.

The cumulative distribution function associated with $f_D$ is defined as:
\begin{equation*}
	F_D(x)
	=
	\int_{-\infty}^{x}
	f_D(s)\,ds.
\end{equation*}

On an interval $I_i=[z_{i-1},z_i]$, it can be written as:
\begin{equation*}
	F_D(x)
	=
	\sum_{k=1}^{i-1}W_k
	+
	W_i
	\frac{x-z_{i-1}}{z_i-z_{i-1}},
	\qquad
	x\in[z_{i-1},z_i].
\end{equation*}

Let:
\begin{equation*}
	S_i
	=
	\sum_{k=1}^{i}W_k
\end{equation*}
denote the cumulative sum of the weights. Then $F_D$ is a continuous distribution function, strictly increasing on intervals for which $W_i>0$, and satisfies:
\begin{equation*}
	F_D(z_i)
	=
	S_i.
\end{equation*}

To generate an ordered pseudo-observation, we draw:
\begin{equation*}
	u_i
	\sim
	U(0,1),
\end{equation*}
and then compute:
\begin{equation*}
	x_i^*
	=
	F_D^{-1}(u_i).
\end{equation*}

In practice, the index $k$ is identified such that:
\begin{equation*}
	S_{k-1}
	<
	u_i
	\leq
	S_k,
\end{equation*}
with:
\begin{equation*}
	S_0=0.
\end{equation*}

The pseudo-observation is then obtained by interpolation within interval
$I_k$:
\begin{equation*}
	x_i^*
	=
	z_{k-1}
	+
	\frac{u_i-S_{k-1}}{W_k}
	(z_k-z_{k-1}).
\end{equation*}

This relationship corresponds exactly to the inversion of the cumulative distribution function $F_D$ on the selected interval.

To stabilize the extremes, corrected uniform draws may also be used:
\begin{equation*}
	u_i
	\sim
	U(a_i,b_i),
\end{equation*}
with:
\begin{equation*}
	a_i
	=
	\max
	\left(
	0.001,
	\frac{i-0.75}{n+0.5}
	\right),
	\qquad
	b_i
	=
	\min
	\left(
	0.999,
	\frac{i-0.25}{n+0.5}
	\right).
\end{equation*}

This correction avoids values that are excessively close to $0$ or $1$, which may generate excessively extreme pseudo-observations.

The preceding generation procedure produces an ordered collection of pseudo-observations:
\begin{equation*}
	x_{(1)}^*,
	x_{(2)}^*,
	\ldots,
	x_{(n)}^*.
\end{equation*}

However, a time series is not merely a marginal distribution; it also possesses a temporal organization. To reconstruct this organization, the method uses the ranks of the original series. The pseudo-time series is defined as:
\begin{equation*}
	X_t^*
	=
	x_{(r_t)}^*,
	\qquad
	t=1,\ldots,n.
\end{equation*}

Thus, if observation $x_t$ occupied rank $r_t$ in the original series, the
bootstrap value at the same time point is assigned the pseudo-observation
corresponding to the same rank.
\begin{algorithm}[t]
	\caption{Computational Dirichlet-Based Quantile Bootstrap}
	\label{alg:dpqbootstrap}
	
	\begin{algorithmic}[1]
		
		\Require Observed time series
		$X=(x_1,\ldots,x_n)$,
		number of bootstrap replications $B$,
		and concentration parameter $\alpha$
		
		\Ensure Bootstrap pseudo-series
		$X^{*(1)},\ldots,X^{*(B)}$
		and bootstrap statistics
		$T^{*(1)},\ldots,T^{*(B)}$
		
		\State \textbf{Step 1: Order the series}
		\Statex \hspace{\algorithmicindent}
		$x_{(1)}\leq x_{(2)}\leq\cdots\leq x_{(n)}$
		
		\State \textbf{Step 2: Compute the ranks}
		\For{$t=1,\ldots,n$}
		\State $r_t\gets\operatorname{rank}(x_t)$
		\EndFor
		
		\State \textbf{Step 3: Construct the interval boundaries}
		\For{$i=1,\ldots,n-1$}
		\State
		$z_i\gets
		\dfrac{x_{(i)}+x_{(i+1)}}{2}$
		\EndFor
		
		\State
		$z_0\gets
		x_{(1)}
		-
		\dfrac{x_{(2)}-x_{(1)}}{2}$
		
		\State
		$z_n\gets
		x_{(n)}
		+
		\dfrac{x_{(n)}-x_{(n-1)}}{2}$
		
		\For{$b=1,\ldots,B$}
		
		\State \textbf{Step 4: Draw Dirichlet weights}
		\State
		$W^{(b)}
		=
		(W_1^{(b)},\ldots,W_n^{(b)})
		\sim
		\operatorname{Dirichlet}
		(\alpha,\ldots,\alpha)$
		
		\State \textbf{Step 5: Construct the cumulative distribution function}
		\State Define the density
		\[
		f_D^{(b)}(x)
		=
		\sum_{i=1}^{n}
		\frac{W_i^{(b)}}{z_i-z_{i-1}}
		\mathbf{1}_{[z_{i-1},z_i]}(x)
		\]
		
		\State Construct the corresponding cumulative distribution function
		$F_D^{(b)}$
		
		\State \textbf{Step 6: Generate pseudo-observations}
		\For{$i=1,\ldots,n$}
		
		\State Draw
		$u_i\sim U(0,1)$
		
		\State Generate
		\[
		x_{(i)}^{*(b)}
		\gets
		\left(F_D^{(b)}\right)^{-1}(u_i)
		\]
		
		\EndFor
		
		\State \textbf{Step 7: Reconstruct the pseudo-time series}
		\For{$t=1,\ldots,n$}
		\State
		$x_t^{*(b)}
		\gets
		x_{(r_t)}^{*(b)}$
		\EndFor
		
		\State
		$X^{*(b)}
		\gets
		(x_1^{*(b)},\ldots,x_n^{*(b)})$
		
		\State \textbf{Step 8: Compute the bootstrap statistic}
		\State
		$T^{*(b)}
		\gets
		T\left(X^{*(b)}\right)$
		
		\EndFor
		
		\State \textbf{Output:} $\{X^{*(b)},T^{*(b)}\}_{b=1}^{B}$

	\end{algorithmic}
\end{algorithm}

This step preserves the relative structure of the series. The temporal positions associated with low, intermediate, and high values are maintained. The method therefore does not resample the observations independently; instead, it reconstructs a pseudo-series while preserving the relative ordering imposed by the original ranks. The procedure is repeated $B$ times. For each replication $b$, a new weight
vector is drawn:
\begin{equation*}
	W^{(b)}
	\sim
	\operatorname{Dirichlet}
	(\alpha,\ldots,\alpha),
\end{equation*}
After which a pseudo-series is generated:
\begin{equation*}
	X^{*(b)}
	=
	\left(
	X_1^{*(b)},
	X_2^{*(b)},
	\ldots,
	X_n^{*(b)}
	\right).
\end{equation*}

The collection of pseudo-series is therefore:
\begin{equation*}
	X^{*(1)},
	X^{*(2)},
	\ldots,
	X^{*(B)}.
\end{equation*}

From these series, a bootstrap mean series can be calculated:
\begin{equation*}
	\bar{X}_t^*
	=
	\frac{1}{B}
	\sum_{b=1}^{B}
	X_t^{*(b)}.
\end{equation*}

Uncertainty bands can also be calculated at each time point $t$. The interquartile interval is given by:
\begin{equation*}
	[Q_{25\%,t},Q_{75\%,t}],
\end{equation*}
where:
\begin{equation*}
	Q_{25\%,t}
	=
	\operatorname{Quantile}_{0.25}
	\left(
	X_t^{*(1)},\ldots,X_t^{*(B)}
	\right),
	\qquad
	Q_{75\%,t}
	=
	\operatorname{Quantile}_{0.75}
	\left(
	X_t^{*(1)},\ldots,X_t^{*(B)}
	\right).
\end{equation*}

A wider interval can also be constructed:
\begin{equation*}
	[Q_{2.5\%,t},Q_{97.5\%,t}].
\end{equation*}

The interval $[Q_{25\%},Q_{75\%}]$ measures central uncertainty, whereas the interval $[Q_{2.5\%},Q_{97.5\%}]$ represents a broader range of uncertainty.

For clarity and reproducibility, the complete computational procedure underlying DPQBootstrap is summarized in Algorithm~\ref{alg:dpqbootstrap}. Starting from the observed time series, the algorithm orders the observations and computes their original ranks, constructs the interval boundaries, and generates Dirichlet probability weights for each bootstrap replication. These weights define a continuous piecewise distribution from which pseudo-observations are generated through quantile inversion. The pseudo-observations are subsequently mapped back to their temporal positions according to the original ranks, yielding a bootstrap pseudo-series from which the statistic of interest can be computed.

\subsection{Difference Between the Simplified Maximum Entropy Bootstrap and DPQBootstrap}

The central difference between the simplified Maximum Entropy Bootstrap and the DPQBootstrap lies in how probability masses are assigned to the intervals constructed around the ordered observations. In the simplified Maximum Entropy Bootstrap, each interval receives a fixed probability mass equal to $1/n$. In DPQBootstrap, this fixed mass is replaced by a random weight $W_i$ drawn from a Dirichlet distribution. Thus, both approaches rely on a piecewise-constant density, but they differ in the nature of the weights associated with the intervals.

Let an observed time series be given by:
\begin{equation*}
	X=(x_1,x_2,\ldots,x_n).
\end{equation*}

Let:
\begin{equation*}
	x_{(1)}\leq x_{(2)}\leq\cdots\leq x_{(n)}
\end{equation*}
denote the ordered observations.

From these values, intervals are constructed as:
\begin{equation*}
	I_i=[z_{i-1},z_i],
	\qquad
	i=1,\ldots,n.
\end{equation*}

In the simplified Maximum Entropy Bootstrap, each interval is assigned the same probability mass:
\begin{equation*}
	P(X^{*} \in I_i)=\frac{1}{n}.
\end{equation*}

The density on interval $I_i$ is therefore given by:
\begin{equation*}
	f^{*}_{i}
	=
	\frac{1}{n(z_i-z_{i-1})}.
\end{equation*}

Thus, the bootstrap density can be written as:
\begin{equation*}
	f^{*}(x)
	=
	\sum_{i=1}^{n}
	\frac{1}{n(z_i-z_{i-1})}
	\mathbf{1}_{[z_{i-1},z_i]}(x).
\end{equation*}

This formulation implicitly imposes:
\begin{equation*}
	W_{i}=\frac{1}{n},
	\qquad
	i=1,\ldots,n.
\end{equation*}

In other words, all intervals receive the same probability mass, even though their density values may differ. Indeed, the height of the density depends on the interval width $z_i-z_{i-1}$. A narrow interval has a higher density, whereas a wider interval has a lower density.

In DPQBootstrap, the same interval construction is retained:
\begin{equation*}
	I_i=[z_{i-1},z_i],
	\qquad
	i=1,\ldots,n.
\end{equation*}

However, instead of fixing the probability masses at $1/n$, a vector of random weights is introduced:
\begin{equation*}
	(W_1,W_2,\ldots,W_n)
	\sim
	\operatorname{Dirichlet}
	(\alpha,\ldots,\alpha).
\end{equation*}

Each interval is then assigned a random probability mass:
\begin{equation*}
	P(X^{*} \in I_i)=W_i.
\end{equation*}

The associated density becomes:
\begin{equation*}
	f_D(x)
	=
	\sum_{i=1}^{n}
\frac{W_{i}}{z_i-z_{i-1}}
\mathbf{1}_{[z_{i-1},z_i]}(x).
\end{equation*}

Thus, the probability mass assigned to each interval is no longer deterministic. It varies from one bootstrap replication to another. This variation introduces additional uncertainty into the generating distribution of the pseudo-observations. The method is therefore probabilistic and has a natural Bayesian interpretation.

The generation of pseudo-observations subsequently relies on an idea common to both methods: inversion of the cumulative distribution function. In the simplified Maximum Entropy Bootstrap, the cumulative distribution function is constructed from the density $f_{D}$, and pseudo-observations are generated as:
\begin{equation*}
	x_i^*
	=
	(F^{*})^{-1}(u_i),
	\qquad
	u_i\sim U(0,1).
\end{equation*}

In DPQBootstrap, the cumulative distribution function is constructed from the density $f_D$, and pseudo-observations are generated as:
\begin{equation*}
	x_i^*
	=
	F_D^{-1}(u_i),
	\qquad
	u_i\sim U(0,1).
\end{equation*}

Therefore, the difference between the two methods does not lie in the principle of inverse quantile transformation itself, but rather in the cumulative distribution function used for generation. In the simplified Maximum Entropy Bootstrap, $F_{\mathrm{MEB}}$ is derived from a density in which all interval probability masses are fixed at $1/n$. In DPQBootstrap, $F_D$ is derived from a density in which the interval probability masses are random and follow a Dirichlet distribution.

A particular case helps clarify the relationship between the two methods. If, in DPQBootstrap, all weights were fixed at:
\begin{equation*}
	W_i=\frac{1}{n},
	\qquad
	i=1,\ldots,n,
\end{equation*}
then the density would become:
\begin{equation*}
	f_D(x)
	=
	\sum_{i=1}^{n}
	\frac{1/n}{z_i-z_{i-1}}
	\mathbf{1}_{[z_{i-1},z_i]}(x).
\end{equation*}

Consequently, we would obtain:
\begin{equation*}
	f_D(x)=f^{*}(x).
\end{equation*}

Thus, the simplified Maximum Entropy Bootstrap can be viewed as a particular case of DPQBootstrap in which all Dirichlet weights are replaced by the fixed, equal weight $1/n$. In contrast, under DPQBootstrap, the weights are randomly drawn:
\begin{equation*}
	(W_1,\ldots,W_n)
	\sim
	\operatorname{Dirichlet}
	(\alpha,\ldots,\alpha),
\end{equation*}
which implies that the density changes across bootstrap replications.

\section{Simulation Study}

\subsection{Simulation Study and Considered Scenarios}

To evaluate the performance of the DPQBootstrap package, a simulation study was conducted by comparing the proposed method with the meboot package, which implements the Maximum Entropy Bootstrap. This comparison is not intended to contrast parametric forecasting models, but rather to analyze two resampling methods applied to time-series data. The primary objective is to evaluate each method's ability to generate bootstrap pseudo-series that preserve the essential characteristics of the original series.

From this perspective, performance is not defined in terms of the absolute superiority of one method over the other. Rather, it is assessed through the ability of the generated pseudo-series to preserve several statistical and temporal properties of the observed series, including its distribution, variability, quantiles, extreme values, and part of its temporal dependence structure. The comparison is therefore conducted within a probabilistic resampling framework rather than within a forecasting framework based on ARMA, SARMA, or SARIMA models.

The number of Monte Carlo repetitions is fixed at
\begin{equation*}
	M=100.
\end{equation*}

Thus, each scenario is repeated one hundred times in order to reduce the effect of simulation randomness and obtain more stable results. For each simulated series, the number of bootstrap replications is fixed at
\begin{equation*}
	B=1000.
\end{equation*}

Consequently, at each Monte Carlo repetition, each of the two methods generates one thousand bootstrap pseudo-series. The two approaches compared are the Maximum Entropy Bootstrap, implemented in the meboot package, and the Dirichlet Quantile Bootstrap, implemented in the DPQBootstrap package.

The simulation study is based on eight scenarios designed to represent different time-series structures: nonlinear trend, irregular seasonality, structural break, time-varying variance, asymmetric extreme values, non-Gaussian noise, asymmetric noise, and a short time series. For the first seven scenarios, the sample size is fixed at
\begin{equation*}
	n=80.
\end{equation*}

The eighth scenario corresponds to the case of a short time series, with
\begin{equation*}
	n=30.
\end{equation*}

The first scenario considers a nonlinear trend:
\begin{equation*}
	X_t
	=
	20
	+
	0.05t
	+
	0.002t^2
	+
	\varepsilon_t.
\end{equation*}

This scenario allows evaluating the methods' ability to reproduce a progressive, nonlinear evolution.

The second scenario introduces irregular seasonality:
\begin{equation*}
	X_t
	=
	20
	+
	3\sin\left(
	\frac{2\pi t}{12}
	\right)
	+
	1.5\cos\left(
	\frac{2\pi t}{6}
	\right)
	+
	\varepsilon_t.
\end{equation*}

This scenario allows examination of the behavior of the methods when the series exhibits periodic fluctuations.

The third scenario corresponds to a structural break:
\begin{equation*}
	X_t
	=
	\begin{cases}
		20+\varepsilon_t,
		& t\leq T/2,\\
		28+\varepsilon_t,
		& t>T/2.
	\end{cases}
\end{equation*}

It is used to assess the methods' ability to handle an abrupt change in level.

The fourth scenario introduces time-varying variance:
\begin{equation*}
	X_t
	=
	20+\varepsilon_t,
\end{equation*}
where
\begin{equation*}
	\varepsilon_t
	\sim
	\mathcal{N}(0,1),
	\qquad
	t\leq T/2,
\end{equation*}
and
\begin{equation*}
	\varepsilon_t
	\sim
	\mathcal{N}(0,4),
	\qquad
	t>T/2.
\end{equation*}

This scenario allows evaluation of the methods' ability to maintain a non-constant level of uncertainty over time.

The fifth scenario introduces positive extreme values through contaminated noise:
\begin{equation*}
	\varepsilon_t
	\sim
	0.85\mathcal{N}(0,1)
	+
	0.15\mathcal{N}(6,4).
\end{equation*}

It is used to examine the robustness of the methods in the presence of asymmetric outliers.

The sixth scenario is based on non-Gaussian heavy-tailed noise:
\begin{equation*}
	\varepsilon_t
	\sim
	t_3.
\end{equation*}

The seventh scenario uses asymmetric noise defined by
\begin{equation*}
	\varepsilon_t
	=
	Z_t-\mathbb{E}(Z_t),
	\qquad
	Z_t
	\sim
	\Gamma(2,1).
\end{equation*}

Finally, the eighth scenario considers a short time series. This case is particularly important because resampling methods are often used when the available sample size is limited.

For each simulated series, six statistics are considered:
\begin{equation*}
	\bar{X},
	\qquad
	\operatorname{median}(X),
	\qquad
	\operatorname{sd}(X),
	\qquad
	Q_{90\%}(X),
	\qquad
	\max(X),
	\qquad
	\rho_1.
\end{equation*}

These statistics represent, respectively, the mean, median, standard deviation, the $90\%$ quantile, the maximum, and the lag-1 autocorrelation. They enable evaluation of several important dimensions of the series: its average level, central location, dispersion, upper values, extreme values, and first-order temporal dependence.

For each statistic, bootstrap intervals are constructed at the levels
\begin{equation*}
	95\%,
	\qquad
	90\%,
	\qquad
	80\%,
	\qquad
	75\%.
\end{equation*}

The primary comparison criterion is the Interval Score. For a bootstrap interval
\begin{equation*}
	[L,U],
\end{equation*}
a reference statistic $T$, and an error level $\alpha$, it is defined as
\begin{equation*}
	\operatorname{IS}_{\alpha}
	=
	(U-L)
	+
	\frac{2}{\alpha}
	(L-T)
	\mathbf{1}_{\{T<L\}}
	+
	\frac{2}{\alpha}
	(T-U)
	\mathbf{1}_{\{T>U\}}.
\end{equation*}

This measure combines two complementary aspects: the width of the interval and its ability to contain the reference statistic. A lower Interval Score therefore indicates a better trade-off between precision and coverage. It penalizes both excessively wide intervals and intervals that fail to contain the statistic under consideration.

Overall, the comparison is based on
\begin{equation*}
	8\times6\times4=192
\end{equation*}
evaluation situations, corresponding to the eight scenarios, the six statistics considered, and the four interval levels.
\subsection{Results of the Comparison Between DPQBootstrap and meboot}

The overall comparison between DPQBootstrap and the Maximum Entropy Bootstrap implemented in the meboot package is summarized in Table~\ref{tab:overall_results}. DPQBootstrap obtained 106 wins, compared with 86 wins for meboot. In percentage terms, this corresponds to $55.21\%$ of the comparison cases in favor of DPQBootstrap and $44.79\%$ in favor of meboot.

\begin{table}[htbp]
	\centering
	\caption{Overall results of the comparison between DPQBootstrap and the Maximum Entropy Bootstrap according to the Interval Score.}
	\label{tab:overall_results}
	\begin{tabular}{lcc}
		\hline
		Method & Number of wins & Percentage \\
		\hline
		DPQBootstrap & 106 & 55.21\% \\
		Maximum Entropy Bootstrap (meboot) & 86 & 44.79\% \\
		\hline
	\end{tabular}
\end{table}

As shown in Table~\ref{tab:overall_results}, the overall result does not imply that DPQBootstrap systematically outperforms meboot in all situations. Rather, it indicates that, within the present simulation study, DPQBootstrap more frequently achieved a lower Interval Score. In other words, it more frequently provided a better trade-off between the width of the bootstrap intervals and their ability to contain the reference statistics.

A scenario-specific analysis is presented in Table~\ref{tab:scenario_results}. This analysis provides a more detailed view of the conditions under which each resampling method obtained the lower Interval Score.

\begin{table}[htbp]
	\centering
	\caption{Number of wins by simulation scenario according to the Interval Score.}
	\label{tab:scenario_results}
	\begin{tabular}{clccc}
		\hline
		No. & Scenario & DPQBootstrap & meboot & Result \\
		\hline
		1 & Nonlinear trend & 8 & 16 & meboot \\
		2 & Irregular seasonality & 15 & 9 & DPQBootstrap \\
		3 & Structural break & 12 & 12 & Tie \\
		4 & Time-varying variance & 15 & 9 & DPQBootstrap \\
		5 & Asymmetric outliers & 12 & 12 & Tie \\
		6 & Non-Gaussian noise & 16 & 8 & DPQBootstrap \\
		7 & Gamma asymmetric noise & 12 & 12 & Tie \\
		8 & Short time series & 16 & 8 & DPQBootstrap \\
		\hline
	\end{tabular}
\end{table}

As reported in Table~\ref{tab:scenario_results}, these results show that meboot performs better in the nonlinear trend scenario. In contrast, DPQBootstrap performs better in scenarios with irregular seasonality, time-varying variance, heavy-tailed non-Gaussian noise, and short time series. In the structural break, asymmetric outlier, and asymmetric Gamma noise scenarios, the two methods exhibit equivalent performance.

The advantage of DPQBootstrap therefore appears primarily when the series presents an irregular structure, time-varying uncertainty, a non-Gaussian distribution, or a limited number of observations. These situations correspond to contexts in which resampling methods may encounter difficulties in simultaneously preserving quantiles, variability, and certain temporal characteristics.

Overall, the results suggest that DPQBootstrap constitutes a competitive alternative to the Maximum Entropy Bootstrap. Its advantage should not be interpreted as universal superiority, but rather as a greater likelihood, within the simulated settings considered in this study, of preserving the statistical and temporal characteristics of the generated time series. DPQBootstrap therefore appears to be a relevant probabilistic resampling method for time-series analysis, particularly when the objective is to generate pseudo-series that retain the empirical structure of the original series.

\section{Empirical Application to Monthly Data from 2013--2017}

The data used in this empirical application comprise monthly observations from January 2013 to December 2017. They describe the evolution in the number of diabetic patients registered at Mohamed Boudiaf Hospital in Ouargla. The series therefore contains 60 observations distributed over five years, as presented in Table~\ref{tab:diabetes_data}.

The objective of this application is to evaluate the contribution of DPQBootstrap within a Bayesian forecasting framework by comparing standard Bayesian ARMA models with their counterparts augmented through DPQBootstrap resampling.

\begin{table}[htbp]
	\centering
	\caption{Monthly numbers of diabetic patients registered at Mohamed Boudiaf Hospital in Ouargla, 2013--2017.}
	\label{tab:diabetes_data}
	\begin{tabular}{lrrrrr}
		\hline
		Month & 2013 & 2014 & 2015 & 2016 & 2017 \\
		\hline
		January   & 557 & 495 & 501 & 671 & 860 \\
		February  & 359 & 428 & 600 & 556 & 1083 \\
		March     & 402 & 525 & 200 & 754 & 777 \\
		April     & 548 & 592 & 302 & 761 & 800 \\
		May       & 547 & 609 & 700 & 814 & 939 \\
		June      & 546 & 701 & 560 & 917 & 461 \\
		July      & 414 & 278 & 200 & 480 & 822 \\
		August    & 243 & 330 & 600 & 659 & 922 \\
		September & 424 & 496 & 411 & 450 & 717 \\
		October   & 466 & 459 & 800 & 928 & 1008 \\
		November  & 444 & 429 & 788 & 860 & 893 \\
		December  & 534 & 581 & 700 & 874 & 400 \\
		\hline
	\end{tabular}
\end{table}

As shown in Table~\ref{tab:diabetes_data}, the observed series provides monthly information on the number of diabetic patients over a five-year period.

Four Bayesian forecasting approaches are compared. The first consists of a Bayesian ARMA$(1,1)$ model estimated directly from the training series. The second consists of a Bayesian ARMA$(2,1)$ model. The third combines DPQBootstrap with a Bayesian ARMA$(1,1)$ model, whereas the fourth combines DPQBootstrap with a Bayesian ARMA$(2,1)$ model.

For the approaches incorporating DPQBootstrap, multiple pseudo-time series are first generated from the training series. A Bayesian ARMA model is subsequently estimated from these bootstrapped trajectories to generate a predictive distribution.

The role of DPQBootstrap is therefore to generate a collection of possible forecasting trajectories through resampling. Let the observed series be denoted by:
\begin{equation*}
	X=(x_1,x_2,\ldots,x_n).
\end{equation*}

DPQBootstrap then generates $B$ pseudo-time series:
\begin{equation*}
	X^{*(1)},X^{*(2)},\ldots,X^{*(B)}.
\end{equation*}

Each pseudo-series can subsequently be used as the basis for a forecasting model. In the case of a Bayesian ARMA model, this procedure produces a collection of forecasts at horizon $h$:
\begin{equation*}
	\widehat{X}_{t+h}^{*(1)},
	\widehat{X}_{t+h}^{*(2)},
	\ldots,
	\widehat{X}_{t+h}^{*(B)}.
\end{equation*}

This collection of forecasts allows computation of a mean forecast and predictive intervals at different nominal levels.

In this application, predictive intervals are evaluated at four nominal levels:
\begin{equation*}
	95\%,\qquad 90\%,\qquad 80\%,\qquad 75\%.
\end{equation*}

The corresponding predictive intervals are defined using the empirical quantiles of the predictive distribution. The $95\%$ predictive interval is:
\begin{equation*}
	\left[
	Q_{2.5\%},
	Q_{97.5\%}
	\right].
\end{equation*}

The $90\%$ predictive interval is:
\begin{equation*}
	\left[
	Q_{5\%},
	Q_{95\%}
	\right].
\end{equation*}

The $80\%$ predictive interval is:
\begin{equation*}
	\left[
	Q_{10\%},
	Q_{90\%}
	\right].
\end{equation*}

Finally, the $75\%$ predictive interval is:
\begin{equation*}
	\left[
	Q_{12.5\%},
	Q_{87.5\%}
	\right].
\end{equation*}

The four forecasting approaches are compared using several evaluation criteria. Point forecast errors are assessed using the Mean Absolute Error (MAE), the Root Mean Squared Error (RMSE), and the Mean Absolute Percentage Error (MAPE). The quality of the predictive intervals is evaluated using empirical coverage, average width, and the Interval Score. Coverage represents the proportion of observed values that fall within the predictive interval. Width measures the degree of dispersion of the interval. The Interval Score combines these two dimensions by penalizing both excessively wide intervals and observations that fall outside the interval.

\begin{figure}[htbp]
	\centering
	\includegraphics[width=\textwidth]{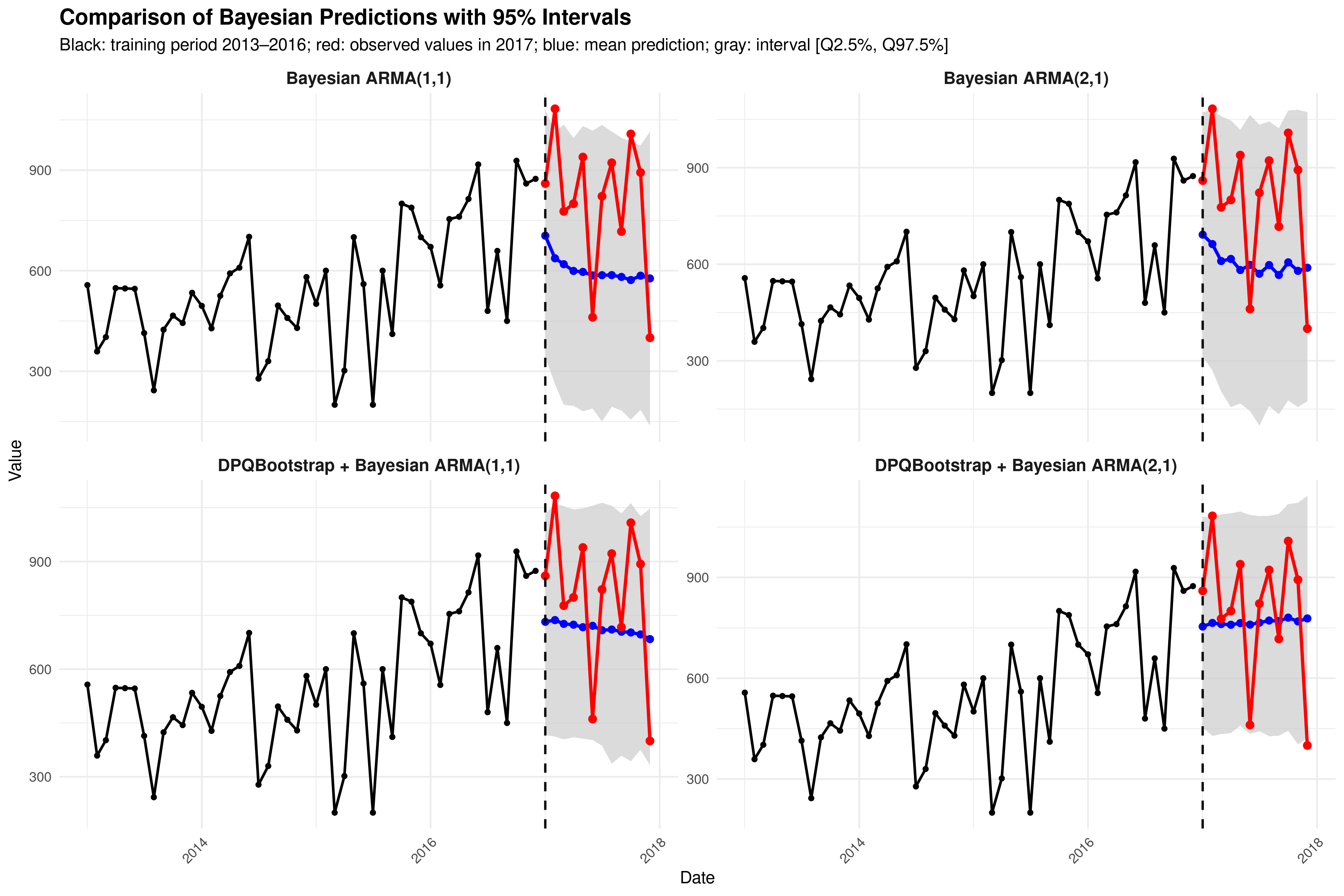}
	\caption{Comparison of out-of-sample forecasts obtained from Bayesian ARMA models with and without DPQBootstrap. The upper panels correspond to Bayesian ARMA$(1,1)$ and Bayesian ARMA$(2,1)$, whereas the lower panels correspond to DPQBootstrap combined with the respective Bayesian ARMA models. The black line denotes the observed training series, the vertical dashed line marks the forecasting origin, the red line denotes the observed out-of-sample values, the blue line represents the point forecasts, and the shaded regions show the $95\%$ predictive intervals.}\caption{Comparison of Bayesian forecasts with and without DPQBootstrap.}
	\label{fig:forecast_comparison}
\end{figure}

Figure~\ref{fig:forecast_comparison} presents a visual comparison of the Bayesian forecasts obtained with and without DPQBootstrap. The figure displays the observed series together with the forecasts and their corresponding $95\%$ predictive intervals for the four approaches considered.

Table~\ref{tab:interval_95} presents the evaluation results for the four methods using $95\%$ predictive intervals.

\begin{table}[htbp]
	\centering
	\caption{Evaluation of Bayesian ARMA and DPQBootstrap--Bayesian ARMA models according to the Interval Score at the $95\%$ level.}
	\label{tab:interval_95}
	\begin{tabular}{lrrrrrr}
		\hline
		Method & MAE & RMSE & MAPE & Coverage & Width & Interval Score \\
		\hline
		Bayesian ARMA $(1,1)$
		& 250.29 & 272.85 & 30.79\% & 83.33\% & 817.70 & 1029.55 \\
		
		Bayesian ARMA $(2,1)$
		& 248.27 & 266.85 & 30.91\% & 91.67\% & 897.25 & 934.58 \\
		
		DPQBootstrap + Bayesian ARMA $(1,1)$
		& 183.94 & 210.57 & 25.40\% & 91.67\% & 669.33 & 751.85 \\
		
		DPQBootstrap + Bayesian ARMA $(2,1)$
		& 162.56 & 198.58 & 24.50\% & 83.33\% & 663.86 & 731.84 \\
		\hline
	\end{tabular}
\end{table}

As shown in Table~\ref{tab:interval_95}, for the $95\%$ interval, the DPQBootstrap + Bayesian ARMA $(2,1)$ approach provides the best overall results. It achieves the lowest values of MAE, RMSE, MAPE, Width, and Interval Score. The Bayesian ARMA $(2,1)$ model provides higher empirical coverage, but at the cost of a substantially wider interval. The inclusion of DPQBootstrap therefore improves point forecast accuracy and narrows the predictive intervals in this comparison.

Table~\ref{tab:interval_90} presents the results obtained for the $90\%$ predictive interval.

\begin{table}[htbp]
	\centering
	\caption{Evaluation of Bayesian ARMA and DPQBootstrap--Bayesian ARMA models according to the Interval Score at the $90\%$ level.}
	\label{tab:interval_90}
	\begin{tabular}{lrrrrrr}
		\hline
		Method & MAE & RMSE & MAPE & Coverage & Width & Interval Score \\
		\hline
		Bayesian ARMA $(1,1)$
		& 250.29 & 272.85 & 30.79\% & 83.33\% & 680.25 & 1035.69 \\
		
		Bayesian ARMA $(2,1)$
		& 248.27 & 266.85 & 30.91\% & 83.33\% & 737.33 & 915.81 \\
		
		DPQBootstrap + Bayesian ARMA $(1,1)$
		& 183.94 & 210.57 & 25.40\% & 83.33\% & 561.55 & 709.46 \\
		
		DPQBootstrap + Bayesian ARMA $(2,1)$
		& 162.56 & 198.58 & 24.50\% & 75.00\% & 552.03 & 853.88 \\
		\hline
	\end{tabular}
\end{table}

For the $90\%$ interval, Table~\ref{tab:interval_90} shows that DPQBootstrap + Bayesian ARMA $(1,1)$ achieves the lowest Interval Score. This indicates that, according to this criterion, its predictive interval provides the most favorable balance between coverage and interval width. In contrast, DPQBootstrap + Bayesian ARMA $(2,1)$ remains the best-performing method in terms of point forecast errors, achieving the lowest MAE, RMSE, and MAPE.

The results obtained for the $80\%$ predictive interval are reported in Table~\ref{tab:interval_80}.

\begin{table}[htbp]
	\centering
	\caption{Evaluation of Bayesian ARMA and DPQBootstrap--Bayesian ARMA models according to the Interval Score at the $80\%$ level.}
	\label{tab:interval_80}
	\begin{tabular}{lrrrrrr}
		\hline
		Method & MAE & RMSE & MAPE & Coverage & Width & Interval Score \\
		\hline
		Bayesian ARMA $(1,1)$
		& 250.29 & 272.85 & 30.79\% & 58.33\% & 520.96 & 965.33 \\
		
		Bayesian ARMA $(2,1)$
		& 248.27 & 266.85 & 30.91\% & 66.67\% & 555.87 & 865.37 \\
		
		DPQBootstrap + Bayesian ARMA $(1,1)$
		& 183.94 & 210.57 & 25.40\% & 58.33\% & 435.44 & 695.80 \\
		
		DPQBootstrap + Bayesian ARMA $(2,1)$
		& 162.56 & 198.58 & 24.50\% & 66.67\% & 430.96 & 735.89 \\
		\hline
	\end{tabular}
\end{table}

As reported in Table~\ref{tab:interval_80}, both DPQBootstrap approaches continue to outperform the standard Bayesian ARMA models across several criteria. DPQBootstrap + Bayesian ARMA $(1,1)$ achieves the lowest Interval Score, whereas DPQBootstrap + Bayesian ARMA $(2,1)$ retains the lowest point forecast errors and produces the narrowest predictive interval. These results suggest that DPQBootstrap can improve predictive performance when narrower predictive intervals are considered.

Finally, Table~\ref{tab:interval_75} presents the results for the $75\%$ predictive interval.

\begin{table}[htbp]
	\centering
	\caption{Evaluation of Bayesian ARMA and DPQBootstrap--Bayesian ARMA models according to the Interval Score at the $75\%$ level.}
	\label{tab:interval_75}
	\begin{tabular}{lrrrrrr}
		\hline
		Method & MAE & RMSE & MAPE & Coverage & Width & Interval Score \\
		\hline
		Bayesian ARMA $(1,1)$
		& 250.29 & 272.85 & 30.79\% & 58.33\% & 466.48 & 912.95 \\
		
		Bayesian ARMA $(2,1)$
		& 248.27 & 266.85 & 30.91\% & 58.33\% & 498.19 & 847.07 \\
		
		DPQBootstrap + Bayesian ARMA $(1,1)$
		& 183.94 & 210.57 & 25.40\% & 50.00\% & 392.45 & 677.74 \\
		
		DPQBootstrap + Bayesian ARMA $(2,1)$
		& 162.56 & 198.58 & 24.50\% & 66.67\% & 384.56 & 686.33 \\
		\hline
	\end{tabular}
\end{table}

For the $75\%$ interval, Table~\ref{tab:interval_75} shows that the approaches incorporating DPQBootstrap produce the narrowest predictive intervals. DPQBootstrap + Bayesian ARMA $(1,1)$ achieves the lowest Interval Score, whereas DPQBootstrap + Bayesian ARMA $(2,1)$ provides the highest empirical coverage and the lowest point forecast errors. These results support the conclusion that integrating DPQBootstrap can improve probabilistic forecasting performance, particularly when relatively compact predictive intervals are considered.

Overall, the results presented in Tables~\ref{tab:interval_95}--\ref{tab:interval_75} show that the methods incorporating DPQBootstrap consistently achieve lower point forecast errors and narrower predictive intervals than the standard Bayesian ARMA models considered in this application. The DPQBootstrap + Bayesian ARMA $(2,1)$ approach is the strongest method for point forecasting, achieving the lowest MAE, RMSE, and MAPE. In contrast, DPQBootstrap + Bayesian ARMA $(1,1)$ achieves the lowest Interval Score for the $90\%$, $80\%$, and $75\%$ predictive intervals.

In the empirical application considered, these results suggest that DPQBootstrap provides an additional contribution to Bayesian ARMA forecasting by generating improved point forecasts and more compact predictive intervals, as evaluated by the criteria considered.
\section{Discussion}

The results of this study show that DPQBootstrap, as an R package dedicated to probabilistic resampling of time series, constitutes a competitive alternative to the Maximum Entropy Bootstrap implemented in the meboot package. The objective of this comparison was not to demonstrate the universal superiority of DPQBootstrap, but rather to evaluate the ability of the two packages to generate pseudo-time series that preserve the essential characteristics of the original series, including the mean, median, variability, quantiles, extreme values, and first-order autocorrelation. This objective is consistent with the general role of the bootstrap as a statistical resampling tool introduced by Efron (1979) and subsequently extended through a probabilistic and Bayesian perspective with Rubin's (1981) Bayesian bootstrap.

Under the adopted evaluation protocol, each scenario is repeated via Monte Carlo replications, while each method generates bootstrapped pseudo-series. The comparison is based on eight scenarios, six statistics, and four interval levels. The primary evaluation criterion is the Interval Score, which simultaneously assesses the width of the bootstrap intervals and their ability to contain the reference statistic. This criterion is particularly appropriate because it does not reward narrow intervals alone, but also penalizes intervals that fail to cover the true statistic adequately.

The overall results indicate that DPQBootstrap achieves 106 wins, compared with 86 for meboot, corresponding to 55.21\% and 44.79\% of the evaluated cases, respectively. This result suggests that DPQBootstrap more frequently provides a better compromise between interval precision and statistical coverage. It does not imply that DPQBootstrap systematically dominates the meboot package, but rather indicates a slightly stronger overall ability to preserve the statistical characteristics of the simulated time series across several contexts.

The scenario-specific analysis, however, provides a more nuanced interpretation of these findings. DPQBootstrap performs particularly favorably in scenarios involving irregular seasonality, with 15 wins compared with 9 for meboot; changing variance, with 15 wins compared with 9; non-Gaussian heavy-tailed noise, with 16 wins compared with 8; and short time series, also with 16 wins compared with 8. These results indicate that DPQBootstrap may be particularly useful when the series exhibits an irregular structure, varying uncertainty, a non-Gaussian distribution, or a limited sample size. In contrast, meboot performs better in the nonlinear trend scenario, with 16 wins compared with 8 for DPQBootstrap. For the structural break, asymmetric outlier, and asymmetric gamma noise scenarios, the two packages achieve equivalent results, with 12 wins each.

These findings are consistent with the methodological differences between the two approaches. The Maximum Entropy Bootstrap proposed by Vinod and López-de-Lacalle (2009) is based on a maximum-entropy framework adapted to time series. DPQBootstrap, in contrast, introduces random Dirichlet-type weighting and reconstructs the pseudo-series through rank preservation. This construction allows the incorporation of an additional source of probabilistic uncertainty while generating continuous pseudo-observations. It may therefore be particularly relevant when the objective is to preserve the empirical distribution, quantiles, extreme values, and part of the series' relative temporal structure.

Nevertheless, the results should be interpreted with caution. DPQBootstrap should not be presented as a systematic replacement for the Maximum Entropy Bootstrap. Its contribution appears to be particularly relevant for complex, short, irregular, or non-Gaussian time series. For series dominated by a smooth trend, the meboot package may remain more effective. DPQBootstrap should therefore be regarded as a complementary package, particularly suited to the probabilistic resampling of time series when the objective is to preserve the statistical and temporal characteristics of the original series.

Finally, this study suggests an important direction for future research: developing a block-based version of DPQBootstrap to preserve local temporal dependence better. Such an extension would bring the method closer to the block bootstrap approaches proposed by Künsch (1989) and Politis and Romano (1994), while retaining the distinctive features of Dirichlet weighting and rank-based reconstruction.
\section{Conclusion}

This study presented DPQBootstrap, an R package for probabilistic resampling of time series via a Dirichlet quantile bootstrap combined with rank-based reconstruction. The package is already available on CRAN under the name DPQBootstrap, allowing its direct use by the scientific community.

The proposed method enables the generation of continuous pseudo-series while preserving the relative structure of the original series. Dirichlet weights introduce probabilistic uncertainty into the empirical distribution, whereas rank-based reconstruction seeks to preserve the temporal organization of the observed values.

The simulation study showed that DPQBootstrap constitutes a competitive alternative to the Maximum Entropy Bootstrap implemented in the meboot package. Across 192 evaluation settings, DPQBootstrap achieved 106 wins, compared with 86 for meboot, for a 55.21\% advantage in favor of DPQBootstrap. These results indicate a stronger overall ability to preserve certain statistical characteristics of the time series, particularly in scenarios involving irregular seasonality, changing variance, non-Gaussian noise, and short series.

Thus, DPQBootstrap should not be presented as a systematic replacement for existing methods, but rather as a complementary, flexible, and useful approach to resampling complex time series. Future extensions could investigate a block-based version of the method to preserve local temporal dependence better.

\section*{Software Availability and Source Code}

The \texttt{DPQBootstrap} package is available on the Comprehensive R Archive Network (CRAN) under the name \texttt{DPQBootstrap}. The package provides an implementation of the Dirichlet Mixture Quantile Bootstrap with rank-based reconstruction for probabilistic resampling of time series.

The complete source code of the package, including the implementation of the DPQBootstrap algorithm and the functions required to generate bootstrap pseudo-series, is available through the package repository and CRAN. The software can be installed directly in \textsf{R} using:

\begin{verbatim}
	install.packages("DPQBootstrap")
\end{verbatim}

The package documentation provides additional information on the available functions, their arguments, and their practical use.

The code used for the simulation study and the empirical Bayesian ARMA forecasting application presented in this article is also available as a reproducible Kaggle notebook: \url{https://www.kaggle.com/code/boulesnane/the-application-of-dpqbootstrap}

\section*{Declaration of generative AI and AI-assisted technologies in the manuscript preparation process}
During the preparation of this work, the authors used ChatGPT to improve the readability and language of the manuscript, and not for idea generation, study design, or the development of scientific content. After using this tool, the authors reviewed and edited the content as needed and take full responsibility for the content of the published article.

\section*{Declaration of competing interest}
The authors declare that they have no known competing financial interests or personal relationships that could have appeared to influence the work reported in this paper.

\bibliographystyle{apalike}

\section*{Appendix}
\appendix
\section{Proof that $f_D$ Is a Probability Density Function}

On interval $i$, the density is constant:
\begin{equation*}
	f_D(x)
	=
	\frac{W_i}{z_i-z_{i-1}},
	\qquad
	x\in[z_{i-1},z_i].
\end{equation*}

The complete density is therefore given by:
\begin{equation*}
	f_D(x)
	=
	\sum_{i=1}^{n}
	\frac{W_i}{z_i-z_{i-1}}
	\mathbf{1}_{[z_{i-1},z_i]}(x),
	\qquad
	(W_1,\ldots,W_n)
	\sim
	\operatorname{Dirichlet}(\alpha,\ldots,\alpha).
\end{equation*}

To show that this is a valid probability density function, consider the integral of the density over each interval. For interval $[z_{i-1},z_i]$, we have:
\begin{equation*}
	\int_{z_{i-1}}^{z_i}
	\frac{W_i}{z_i-z_{i-1}}
	\,dx
	=
	\frac{W_i}{z_i-z_{i-1}}
	(z_i-z_{i-1})
	=
	W_i.
\end{equation*}

Thus, the integral of the density over each interval is exactly equal to the probability mass $W_i$ assigned to that interval.

Consequently, over all intervals:
\begin{equation*}
	\int_{z_0}^{z_n}
	f_D(x)\,dx
	=
	\sum_{i=1}^{n}W_i
	=
	1.
	\tag{A.1}
\end{equation*}

Therefore, $f_D$ is a valid probability density function.

\section{Proof of the Uniform Distribution as the Maximum-Entropy Solution in the Discrete Case}

Suppose that there are $n$ intervals:
\begin{equation*}
	I_1,\ldots,I_n,
\end{equation*}
and that a total probability mass equal to one is to be distributed among these intervals. Define:
\begin{equation*}
	p_i
	=
	P(X^*\in I_i).
	\tag{A.2}
\end{equation*}

The only constraints imposed are:
\begin{equation*}
	p_i\geq 0,
	\qquad
	\sum_{i=1}^{n}p_i=1.
\end{equation*}

The discrete entropy is defined as:
\begin{equation*}
	H(p)
	=
	-
	\sum_{i=1}^{n}
	p_i\ln(p_i).
\end{equation*}

This entropy is maximized when all probability masses are equal:
\begin{equation*}
	p_1
	=
	p_2
	=
	\cdots
	=
	p_n
	=
	\frac{1}{n}.
\end{equation*}

Therefore:
\begin{equation*}
	p_i
	=
	\frac{1}{n},
	\qquad
	i=1,\ldots,n.
\end{equation*}

This corresponds to the least informative allocation of probability mass among the intervals. In other words, when there is no reason to assign greater importance to one interval than to another, the maximum-entropy solution assigns equal probability mass to every interval.

\section{Derivation of the Density Form of the Maximum Entropy Bootstrap}

The Maximum Entropy Bootstrap seeks a density $f^*$ that introduces as little additional information as possible beyond that contained in the observed data. The starting point is therefore the maximum entropy principle:
\begin{equation*}
	f^*
	=
	\arg\max_f
	\left\{
	-\int f(x)\ln f(x)\,dx
	\right\}.
\end{equation*}

However, this density cannot be chosen arbitrarily. It must remain linked to the time series observations. For this purpose, the observations are first arranged in increasing order:
\begin{equation*}
	x_{(1)}
	\leq
	x_{(2)}
	\leq
	\cdots
	\leq
	x_{(n)}.
\end{equation*}

An interval is then constructed around each observation:
\begin{equation*}
	I_i
	=
	[z_{i-1},z_i].
\end{equation*}

Each interval represents one observation from the sample. Since there are $n$ observations, each observation naturally corresponds to an empirical probability mass of:
\begin{equation*}
	\frac{1}{n}.
\end{equation*}

Therefore, according to Equation~(A.2), each interval is assigned the probability mass \footnote{This equal allocation of probability mass corresponds to the maximum-entropy constraint.}:
\begin{equation*}
	P(X^*\in I_i)
	=
	\frac{1}{n}.
\end{equation*}

Within each interval, the objective is to avoid introducing additional information. The least informative density within an interval is uniform, which is therefore constant. Thus, over the interval
\begin{equation*}
	I_i
	=
	[z_{i-1},z_i],
\end{equation*}
we write:
\begin{equation*}
	f^*(x)
	=
	c_i.
\end{equation*}

This density, however, must assign a total probability mass of $1/n$ over the interval. Therefore:
\begin{equation*}
	\int_{z_{i-1}}^{z_i}
	f^*(x)\,dx
	=
	\frac{1}{n}.
\end{equation*}

Since $f^*(x)=c_i$, we obtain:
\begin{equation*}
	c_i
	(z_i-z_{i-1})
	=
	\frac{1}{n}.
\end{equation*}

Hence:
\begin{equation*}
	c_i
	=
	\frac{1/n}{z_i-z_{i-1}}.
\end{equation*}

Thus, on each interval:
\begin{equation*}
	f^*(x)
	=
	\frac{1/n}{z_i-z_{i-1}},
	\qquad
	x\in[z_{i-1},z_i].
\end{equation*}

Consequently, over all intervals, the density can be written as:
\begin{equation*}
	f^*(x)
	=
	\sum_{i=1}^{n}
	\frac{1/n}{z_i-z_{i-1}}
	\mathbf{1}_{[z_{i-1},z_i]}(x).
\end{equation*}

In DPQBootstrap, this fixed probability a random probability mass replaces mass:
\begin{equation*}
	P(X^*\in I_i)
	=
	W_i,
	\qquad
	(W_1,\ldots,W_n)
	\sim
	\operatorname{Dirichlet}
	(\alpha,\ldots,\alpha).
\end{equation*}

\section{Derivation of the Bayesian Cumulative Distribution Function}

We start from the Dirichlet bootstrap density:
\begin{equation*}
	f_D(x)
	=
	\sum_{i=1}^{n}
	\frac{W_i}{\Delta_i}
	\mathbf{1}_{[z_{i-1},z_i]}(x),
	\tag{A.3}
\end{equation*}
where:
\begin{equation*}
	\Delta_i
	=
	z_i-z_{i-1}.
\end{equation*}

The weights satisfy:
\begin{equation*}
	W_i\geq 0,
	\qquad
	\sum_{i=1}^{n}W_i=1.
\end{equation*}

This means that each interval
\begin{equation*}
	I_i
	=
	[z_{i-1},z_i]
\end{equation*}
is assigned a probability mass $W_i$. Since the density is constant over this interval, we have:
\begin{equation*}
	f_D(x)
	=
	\frac{W_i}{z_i-z_{i-1}},
	\qquad
	x\in[z_{i-1},z_i].
\end{equation*}

The cumulative distribution function is defined as:
\begin{equation*}
	F_D(x)
	=
	P(X^*\leq x)
	=
	\int_{-\infty}^{x}
	f_D(s)\,ds.
\end{equation*}

Since the density is equal to zero outside the overall interval
\begin{equation*}
	[z_0,z_n],
\end{equation*}
for
\begin{equation*}
	x\in[z_{i-1},z_i],
\end{equation*}
the integral up to $x$ contains two components:

\begin{enumerate}
	\item the probability masses associated with all preceding intervals;
	\item a partial probability mass associated with the current interval $i$.
\end{enumerate}

Therefore:
\begin{equation*}
	F_D(x)
	=
	\int_{z_0}^{x}
	f_D(s)\,ds.
\end{equation*}

Since $x$ belongs to interval $i$, the integral can be decomposed as:
\begin{align}
	\int_{z_0}^{x}
	f_D(s)\,ds
	={}&
	\int_{z_0}^{z_1}
	f_D(s)\,ds
	+
	\int_{z_1}^{z_2}
	f_D(s)\,ds
	+\cdots \nonumber\\
	&+
	\int_{z_{i-2}}^{z_{i-1}}
	f_D(s)\,ds
	+
	\int_{z_{i-1}}^{x}
	f_D(s)\,ds.
\end{align}

The first terms correspond to intervals that have already been completely traversed:
\begin{equation*}
	[z_0,z_1],
	[z_1,z_2],
	\ldots,
	[z_{i-2},z_{i-1}].
\end{equation*}

For a preceding interval
\begin{equation*}
	I_k
	=
	[z_{k-1},z_k],
	\qquad
	k<i,
\end{equation*}
the density is:
\begin{equation*}
	f_D(s)
	=
	\frac{W_k}{z_k-z_{k-1}}.
\end{equation*}

Therefore:
\begin{equation*}
	\int_{z_{k-1}}^{z_k}
	f_D(s)\,ds
	=
	\int_{z_{k-1}}^{z_k}
	\frac{W_k}{z_k-z_{k-1}}
	\,ds.
\end{equation*}

Since
\begin{equation*}
	\frac{W_k}{z_k-z_{k-1}}
\end{equation*}
is constant, we obtain:
\begin{equation*}
	\int_{z_{k-1}}^{z_k}
	\frac{W_k}{z_k-z_{k-1}}
	\,ds
	=
	\frac{W_k}{z_k-z_{k-1}}
	\int_{z_{k-1}}^{z_k}
	ds.
\end{equation*}

Now:
\begin{equation*}
	\int_{z_{k-1}}^{z_k}
	ds
	=
	z_k-z_{k-1}.
\end{equation*}

Therefore:
\begin{equation*}
	\int_{z_{k-1}}^{z_k}
	f_D(s)\,ds
	=
	\frac{W_k}{z_k-z_{k-1}}
	(z_k-z_{k-1})
	=
	W_k.
\end{equation*}

Thus, the probability mass accumulated before interval $i$ is:
\begin{equation*}
	\sum_{k=1}^{i-1}
	W_k.
	\tag{A.4}
\end{equation*}

We now consider the remaining part of the current interval:
\begin{equation*}
	[z_{i-1},x].
\end{equation*}

Over this interval, the density is:
\begin{equation*}
	f_D(s)
	=
	\frac{W_i}{z_i-z_{i-1}}.
\end{equation*}

Therefore:
\begin{equation*}
	\int_{z_{i-1}}^{x}
	f_D(s)\,ds
	=
	\int_{z_{i-1}}^{x}
	\frac{W_i}{z_i-z_{i-1}}
	\,ds.
\end{equation*}

Since the density is constant:
\begin{equation*}
	\int_{z_{i-1}}^{x}
	\frac{W_i}{z_i-z_{i-1}}
	\,ds
	=
	\frac{W_i}{z_i-z_{i-1}}
	\int_{z_{i-1}}^{x}
	ds.
\end{equation*}

Since:
\begin{equation*}
	\int_{z_{i-1}}^{x}
	ds
	=
	x-z_{i-1},
\end{equation*}
we obtain:
\begin{equation*}
	\int_{z_{i-1}}^{x}
	f_D(s)\,ds
	=
	\frac{W_i}{z_i-z_{i-1}}
	(x-z_{i-1}).
	\tag{A.5}
\end{equation*}

Combining the two components gives:
\begin{equation*}
	F_D(x)
	=
	\sum_{k=1}^{i-1}W_k
	+
	\frac{W_i}{z_i-z_{i-1}}
	(x-z_{i-1}).
	\tag{A.6}
\end{equation*}

Therefore:
\begin{equation*}
		F_D(x)
		=
		\sum_{k=1}^{i-1}W_k
		+
		W_i
		\frac{x-z_{i-1}}
		{z_i-z_{i-1}},
		\qquad
		x\in[z_{i-1},z_i].
\end{equation*}

This is the required expression for the Bayesian cumulative distribution function.

\section{Generation of Pseudo-Observations by the Inverse Quantile Transformation}

The expression to be derived is:
\begin{equation*}
	x_i^*
	=
	z_{k-1}
	+
	\frac{u_i-S_{k-1}}{W_k}
	(z_k-z_{k-1}),
	\tag{A.7}
\end{equation*}
where:
\begin{equation*}
	S_{k-1}
	=
	\sum_{j=1}^{k-1}W_j.
\end{equation*}

This expression is obtained by inverting the cumulative distribution function $F_D$.

It has already been shown that, for:
\begin{equation*}
	x\in[z_{k-1},z_k],
\end{equation*}
the cumulative distribution function is:
\begin{equation*}
	F_D(x)
	=
	S_{k-1}
	+
	W_k
	\frac{x-z_{k-1}}
	{z_k-z_{k-1}}.
\end{equation*}

To generate a pseudo-observation, we draw:
\begin{equation*}
	u_i
	\sim
	U(0,1),
\end{equation*}
and then determine:
\begin{equation*}
	x_i^*
	=
	F_D^{-1}(u_i).
\end{equation*}

This means that:
\begin{equation*}
	F_D(x_i^*)
	=
	u_i.
\end{equation*}

Therefore, if $u_i$ falls within the cumulative probability interval:
\begin{equation*}
	S_{k-1}
	<
	u_i
	\leq
	S_k,
\end{equation*}
then $x_i^*$ belongs to the interval:
\begin{equation*}
	[z_{k-1},z_k].
\end{equation*}

Substituting $x_i^*$ into the expression for $F_D$ gives:
\begin{equation*}
	u_i
	=
	S_{k-1}
	+
	W_k
	\frac{x_i^*-z_{k-1}}
	{z_k-z_{k-1}}.
\end{equation*}

Subtracting $S_{k-1}$ from both sides yields:
\begin{equation*}
	u_i-S_{k-1}
	=
	W_k
	\frac{x_i^*-z_{k-1}}
	{z_k-z_{k-1}}.
\end{equation*}

Dividing both sides by $W_k$ gives:
\begin{equation*}
	\frac{u_i-S_{k-1}}{W_k}
	=
	\frac{x_i^*-z_{k-1}}
	{z_k-z_{k-1}}.
\end{equation*}

Multiplying both sides by $(z_k-z_{k-1})$ gives:
\begin{equation*}
	\frac{u_i-S_{k-1}}{W_k}
	(z_k-z_{k-1})
	=
	x_i^*-z_{k-1}.
\end{equation*}

Finally, adding $z_{k-1}$ to both sides gives:
\begin{equation*}
	x_i^*
	=
	z_{k-1}
	+
	\frac{u_i-S_{k-1}}{W_k}
	(z_k-z_{k-1}).
\end{equation*}

Therefore:
\begin{equation*}
		x_i^*
		=
		z_{k-1}
		+
		\frac{u_i-S_{k-1}}{W_k}
		(z_k-z_{k-1})
\end{equation*}

This expression means that $u_i$ first determines the interval $k$ through the cumulative weights. The pseudo-observation $x_i^*$ is then positioned within the interval $[z_{k-1},z_k]$ through linear interpolation.

\end{document}